# Resolution of Discrete Excited States in InGaN Multiple Quantum Wells using Degenerate Four Wave Mixing

D.O. Kundys[1,a)], J.-P.R. Wells[1], A.D. Andreev[2], S.A. Hashemizadeh[1], T. Wang[3], P.J. Parbrook[3], A.M. Fox[1], D.J. Mowbray[1] and M.S. Skolnick[1]

[1]*Department of Physics and Astronomy, University of Sheffield, Sheffield S3 7RH, United Kingdom.*
[2] *Advanced Technology Institute, University of Surrey, Guildford, GU2 7XH, United Kingdom.*
[3] *EPSRC National Centre for III-V Technologies, University of Sheffield, Sheffield S1 3JD, United Kingdom.*

We report on two pulse, degenerate four wave mixing (DFWM) measurements on shallow InGaN/GaN multi-quantum wells (MQWs) grown on sapphire substrates. These reveal pulse length limited signal decays. We have found a 10:1 resonant enhancement of the DFWM signal at the excitonic transition frequencies which thereby give a sharp discrimination of the discrete excitonic contributions within the featureless distribution seen in absorption spectra. The exciton resonances have peak positions, which yield good overall agreement with a full **k·P** model calculation for the quantum well energy levels and optical transition matrix elements. InGaN/GaN MQWs generally exhibit strongly inhomogeneously broadened excitation spectra due to indium fluctuation effects; this approach therefore affords a practical method to extract information on the excited excitonic states not available previously.

PACS: 73.21.Fg, 78.47.+p, 78.55.Cr, 42.50.Md  Keywords: InGaN, DFWM, time-resolved, semiconductors.

[a)] corresponding email dmytro.kundys@manchester.ac.uk

## I. INTRODUCTION

Wurtzite GaN based alloys have attracted strong interest for optoelectronic applications, in particular the development of blue and UV light emitting diodes (LEDs) and laser diodes. The introduction of indium is significant due to the concomitant carrier localisation and enhanced radiative quantum efficiency. However the alloy nature leads to significant compositional disorder, which causes major broadening of the optical spectra. The lattice mismatch between InN and GaN also leads to strain induced spontaneous and piezoelectric polarisation fields near the interface of a heterostructure and results in a quantum confined Stark effect (QCSE). This spatially separates the electron and hole in confined systems and reduces the optical transition strength [1].

We report on two-pulse photon-echo experiments on low indium concentration, InGaN MQWs as a function of well thickness and excitation wavelength. Time domain techniques have been extensively employed to study InGaN structures, although for the most part these have concentrated on time resolved photoluminescence studies [2]. Pump-probe techniques have also been used to study carrier relaxation and some of these studies have demonstrated quantum beats due to impulsive excitation of acoustic phonons [3, 4]. Previous degenerate four wave mixing (DFWM) spectroscopy in III-V nitrides has mostly concentrated on GaN epilayers where exciton dephasing times, exciton-exciton, exciton-impurity scattering, electron-phonon and exciton-bi-exciton interactions have been studied [5-8]. Typically, dephasing times of ~ 1-2 ps are measured.

InGaN epilayers and MQWs represent a highly inhomogeneously broadened system and excitonic resonances can be difficult to observe directly in a linear absorption spectrum. Our experiments reveal a pulse length limited, DFWM signal in both epilayers and MQWs. The measured signals are resonantly enhanced at the ground and excited state, heavy-hole exciton absorption frequencies, yielding well-resolved resonances against the suppressed background of continuum states. From the dependence of the DFWM signal upon the intensity of the incident beams, it is concluded that the results may be strictly described in terms of the third order optical susceptibility. We have used the resonant nature of these signals to determine the position of excited states in quantum wells of different thickness yielding good agreement with model calculations.

## II. EXPERIMENTAL DETAILS

The samples consisted of 10 period $In_{0.11}Ga_{0.89}N$/GaN MQWs. In all cases, the barrier was 7.5 nm thick, and the quantum well thicknesses were varied between 1, 4, and 8 nm. All investigated samples were grown on (0001) sapphire double polished substrates by metalorganic chemical vapour deposition (MOCVD). The substrates were initially treated in $H_2$ ambient at 1170 $^{o}$C, followed by the growth of a 25 nm thick low-temperature (550$^{o}$C) GaN buffer layer and a 1.5 µm thick layer of nominally undoped GaN grown at a high temperature. Afterwards, the temperature was lowered to grow a 10 period MQW.  The PL spectra were excited with a He-Cd laser, dispersed in a 0.75m monochromator and detected with a GaAs photomultiplier tube. PLE spectra were obtained using a Xenon lamp with wavelength selectivity/tunability provided by a 0.25m monochromator. Time resolved spectroscopy was performed using a Ti-Sapphire oscillator yielding 150 fs pulses, tunable between 700 nm and 1 µm at a repetition rate of 80 MHz. To excite the interband transitions of InGaN alloys, the oscillator output was frequency doubled providing the required excitation wavelengths between 350 and 450 nm. The DFWM measurements were performed in a standard two pulse, non-co-linear 'forward box', four wave mixing geometry with the phase matched signal detected in the $|2k_2-k_1|$ direction, where $k_1$ and $k_2$ are the wave vectors of the echo generating pulse sequence. The maximum available energy of the 'dipole inverting' $k_2$ pulse was 0.47 nJ with the $k_1$ pulse held at an energy of 0.024nJ. Both beams were focussed using 5 cm focal length $CaF_2$ lenses onto the sample position yielding a spot diameter of close to 50

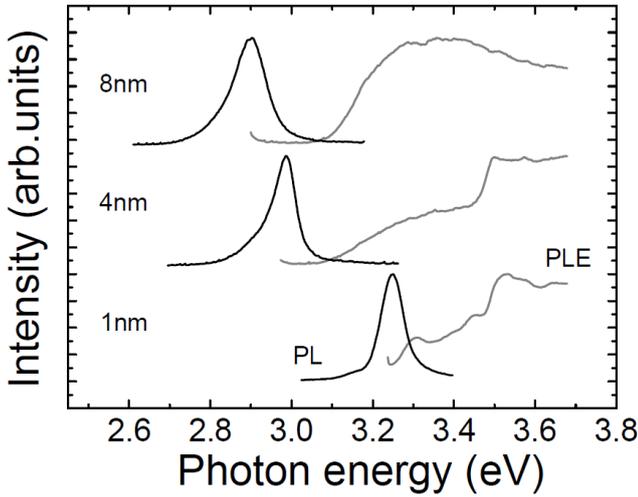

Figure 1. 10 K PL and PLE of the InGaN/GaN multi-quantum well samples.

micron. The signal was detected using a photomultiplier and a phase sensitive detection scheme.

### III. RESULTS AND ANALYSIS

Figure 1 shows the 10 K PL and PLE spectra of the three MQW samples. The PL spectrum for all samples shows a single-dominant peak at 3.25, 2.98 and 2.90 eV for the 1, 2, and 8 nm MQWs respectively. By contrast, the PLE spectrum of the 1 nm thick sample shows one clear peak corresponding to the ground state hh1-e1 excitonic transition in the QW with some evidence for higher energy transitions. The PLE spectra for the 4 and 8 nm thick MQW samples reveal only a broad, structureless increase in absorption with no resolved features which would be indicative of excitonic resonances. For the wider wells, a large shift of the PL from the onset of the quantum well interband absorption is observed. This shift results from a combination of the QCSE and exciton and carrier localisation effects caused by indium disorder fluctuations [9]. It is significant that for the 4 nm and 8 nm thick MQWs samples, the PLE shows no absorption due to states close to the PL energy. This is due to a larger spatial separation of ground state electrons and holes in the wider wells and therefore weaker oscillator strengths of those transitions, which results in their suppression in PLE spectra. The featureless PLE spectra of the 4 and 8 nm MQWs arise due to the contribution of the excited states, which are significantly broadened due to their enhanced sensitivity to both well width *and* electric field fluctuations. This behaviour is in contrast to GaAs quantum wells, which in the absence of electric fields display stronger broadening for narrow wells due to well width fluctuations [10].

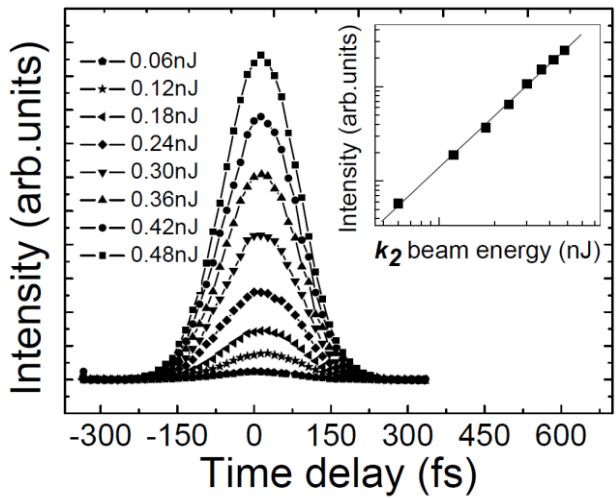

Figure 2. Power dependence of the time-integrated DFWM signal in an 8 nm thick InGaN/GaN MQW for a sample temperature of 5 K. The inset shows the logarithmic plot intensity dependence of the DFWM signal at zero delay (squares), and the linear fit (solid line); the line slope of 2 indicates the quadratic relation.

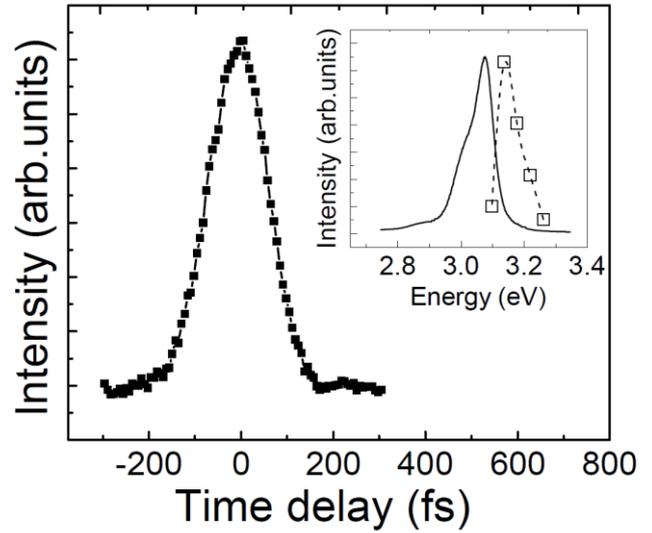

Figure 3. DFWM signal for a 100 nm thick $In_{0.1}Ga_{0.9}N$ epilayer as measured at 5 K. The inset shows the normalised wavelength dependence of DFWM signal (open squares) fitted with a dashed line overlaid as a guide to the eye, and the PL (solid line) spectrum.

In our two pulse DFWM experiments, the observed signal from all samples consists of a very fast, pulse length limited transient. In figure 2, the time integrated (TI) DFWM signal is plotted as a function of the time delay between the generating pulses for the 8 nm thick sample. Transients are recorded for different energies of the $k_2$ pulse with the $k_1$ pulse energy kept constant at 0.024 nJ. The inset to fig. 2 shows the intensity dependence of the DFWM signal at zero delay on a logarithmic scale; the data are fitted well with a line having a slope of 2, showing that we are within the $\chi^{(3)}$ limit. The very fast dephasing we observe is an order of magnitude faster than that observed in GaN epilayers [5-8]. The fast dephasing times observed are most likely attributable to the alloy disorder which leads to a disorder induced dephasing [11] which is dependent on high indium concentrations. An additional contributing factor may be carrier escape tunnelling [12] that can take place due to the built-in electric field in InGaN/GaN MQWs.

We have also performed a study of a 100 nm thick $In_{0.1}Ga_{0.9}N$ epilayer. We found a pulse limited time-integrated DFWM signal that is enhanced at the epilayer band-edge energy (see figure 3). The wavelength dependence of the DFWM signal observed in the epilayer shows a single peak whose intensity increases rapidly from the low energy side with a shoulder on the high energy side reflecting the span of the free exciton splitting influenced by the bandgap bowing in InGaN. This result indicates that DFWM in InGaN systems is enhanced at the excitonic states of the system, a result we show is fully borne out by the measurements on the QWs.

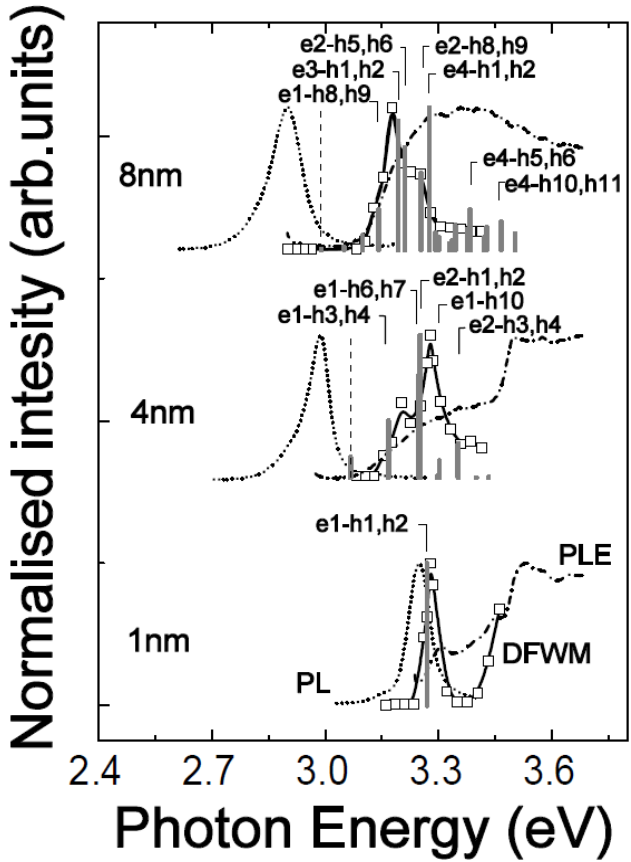

Figure 4. The wavelength dependence of the normalised DFWM signal overlaid on the PL and PLE spectra for the 1, 4 and 8 nm MQWs. The vertical grey lines represent the spectral position (and transition energy) of the interband transitions which have a high oscillator strength. The height of the grey lines is proportional to the absorption strength. The black vertical dashed lines represent *only* the spectral position of the ground state transition in the model calculation, (the transition intensities are negligible).

Figure 4 shows DFWM signals measured at different excitation wavelengths (for the three QW samples discussed above) with a fixed excitation density of 6.4 µJ/cm$^2$. These are plotted together with the normalised PL and PLE spectra for each of the samples. Surprisingly given the broad featureless nature of the PLE for these samples the DFWM results show clear resonances at specific wavelengths. Concentrating on the results for the 1 nm thick sample to begin with, it is immediately evident that the non-linear signal is resonantly enhanced at the lowest e1-hh1 transition of the quantum well. Hence we conclude that the enhancement of the DFWM signal at a specific wavelength indicates the position of the excitonic energy levels in the QW. The DFWM signal is also observed to increase just below the barrier energy which may arise either from a shoulder of strong DFWM signal from the GaN epilayer, or from the presence of a QW state nearly resonant with the barrier band edge in the PLE spectrum.

For the wider wells, the DFWM spectra are more complex with a number of overlapping transitions observed. Such rich structure is suggestive of the observation of excited state transitions within the wells. There is no signal detected for wavelengths close to the PL peak wavelengths in both the 4 and 8 nm thick samples. This is due to the pronounced quantum confined Stark effect (QCSE) in the wider wells leading to weak wave-function overlap for the e1-hh1 transition. The key observation is that clear and distinct resonances can be observed in DFWM even though none are present in the PLE spectra with the exception of the 1 nm sample where the e1-hh1 transition has large oscillator strength.

In order to exemplify that the DFWM signal is shaped by transitions with strong oscillator strength, we employed a multi-band **k·P** model for wurtzite semiconductor to calculate the energy levels and optical transition matrix elements for the samples studied. The **k·P** Hamiltonian employed, including strain and band mixing effects, is described in details in Ref [13, 14]. To calculate the energy levels and optical matrix elements we used plane wave expansion method [14,15]. We used both 8x8 and 4x4 models; the 4x4 model is obtained from the 8x8 model by taking the spin-orbit splitting $\Delta_3$=0. Since for InN and GaN this splitting is very small ($\Delta_3$=4.6 meV, see Table 1), in figure 5 one can see that both models give very similar results (in the 8x8 model we obtain a small additional splitting of the order of $\Delta_3$ for some of the transitions; for the purposes of the present paper this splitting is not important), so we present here the results obtained from the 4x4 model.

TABLE I. Material band structure parameters

| Parameter | GaN | InN | In$_{0.11}$Ga$_{0.89}$N |
|---|---|---|---|
| Lattice constants | | | |
| $c$ (nm) | 0.5185 | 0.5703 | 0.5242 |
| $a$ (nm) | 0.3189 | 0.3545 | 0.3228 |
| Valence band effective mass parameters | | | |
| $A_1$ | −7.21 | −8.21 | −7.32 |
| $A_2$ | −0.44 | −0.68 | −0.47 |
| $A_3$ | 6.68 | 7.57 | 6.78 |
| $A_4$ | −3.46 | −5.23 | −3.65 |
| $A_5$ | −3.40 | −5.11 | −3.59 |
| $A_6$ | −4.90 | −5.96 | −5.02 |
| Deformation potentials[a] | | | |
| $a_c$ (eV) | −4.08 | | −4.08 |
| $D_1$ | 0.7 | | 0.7 |
| $D_2$ | 2.1 | | 2.1 |
| $D_3$ | 1.4 | | 1.4 |
| $D_4$ | −0.7 | | −0.7 |
| Energy parameters | | | |
| $\Delta_1$ (=$\Delta_{cr}$) (eV) | 0.010 | 0.040 | 0.013 |
| $\Delta_2$=$\Delta_3$=$\Delta_{so}/3$ (eV) | 0.005 | 0.0016 | 0.0046 |
| $E_g^{(10\,K)}$ (eV) | 3.51 | 0.78 | 3.07 |
| | | | $b$ (eV)=1.4 |
| Conduction band effective masses | | | |
| $m_e^z$ ($m_0$) | 0.20 | 0.14 | 0.19 |
| $m_e^t$ ($m_0$) | 0.18 | 0.1 | 0.17 |
| Dielectric constant | | | |
| $\varepsilon$ | 10.28 | 14.61 | 10.756 |
| Elastic stiffness constants | | | |
| $C_{11}$ (GPa) | 390 | 223 | 371.6 |
| $C_{12}$ (GPa) | 145 | 115 | 141.7 |
| $C_{13}$ (GPa) | 106 | 92 | 104.5 |
| $C_{33}$ (GPa) | 398 | 224 | 378.9 |
| PZ constants | | | |
| $d_{31}$ (pm/V) | −.6 | −3.5 | −1.81 |
| $d_{33}$ (pm/V) | 3.1 | 7.6 | 3.59 |

[a]Reference 13.

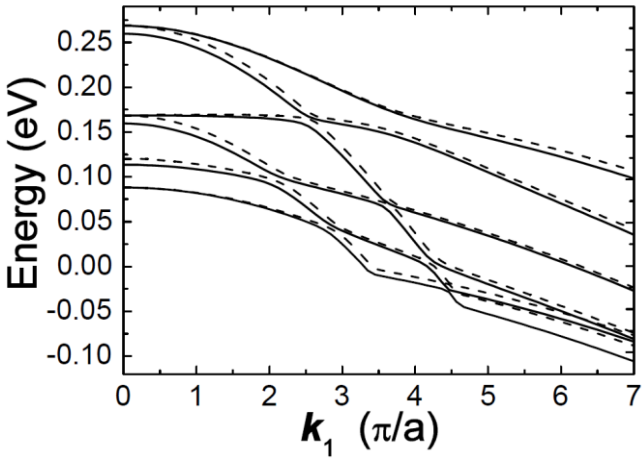

Figure 5. The in-plain dispersion of the valence subbands along <100> direction for 4 nm MQW sample (a=4nm). The calculation results obtained using the 8x8 and 4x4 models are indicated with solid and dashed lines respectively.

Recently, we have shown that the uncertainty in both the elastic and piezoelectric constants lead to a significant inaccuracy in calculated transition energies [16]. The electric field was taken as an adjustable parameter in order to fit the QW ground state transition to our PL data with a reassuring 'Stokes' shift of less then 80 meV. The values employed in our calculations of 1.2, 1 and 0.45 MV/cm for the 1, 4, and 8nm MQWs respectively, are consistent with the findings of ref [17] where a reduction of the electric field was found with increasing well width. Table 1 shows the numerical values of the material constants used in the **k·P** band calculations for unstrained materials, and are taken from ref [18], unless otherwise indicated. The alloy properties of InGaN are obtained by linear interpolation except for the energy gap where the relation from ref [19] is used.

In Figure 4, the vertical grey lines indicate the results of the calculation with the line height representing absorption strengths of the relative interband transitions for each of the samples, while their spectral position represents the transition energy. The calculation of the 1nm QW transition energies shows the presence of only the ground state exciton transition at an energy of 3269 meV, while more a complex structure is observed for the 4 and 8nm QWs. As we can see from Figure 4, the simulation gives good agreement for all studied QW samples proving that the peaks in the DFWM trend indicate the position of high oscillator strength transitions. In the 1 nm QW sample the peak in the DFWM signal corresponds to the ground state e1-hh1 transition and yields a single intense resonance. For the 4 nm QW sample, the calculated transitions are redshifted by approximately 35 meV from the experimental DFWM signal peaks; however the general trend is similar and the relative intensities approximately correct. For the 8 nm thick sample the situation is more complex due to the multitude of possible transitions (fig.4). It is clear that for both the 4 and 8 nm MQWs the DFWM spectra are determined by excited state transitions, the ground state transitions being unobservable due to their very weak oscillator strengths. Good overall agreement is obtained and the calculations indicate that the resonances observed in the DFWM experiments reflect excitonic transitions within the quantum well. The MQWs systems we have investigated have a complex valence band structure [2, 20-21], which are non-isotropic, non-parabolic and semi-degenerate. Thus for the wider wells it is impractical to label all of the transitions. The essential points to be made are: (i) the DFWM spectra indicate the position of excited states in the quantum wells, (ii) the spectra are shaped by strong oscillator strength transitions even though the precise assignments cannot be considered definitive and (iii) the wider wells do not show transitions from low lying quantum well states due to the low oscillator strength of transitions arising from those levels.

Early studies on GaAs QWs [22, 23] revealed that DFWM signals in quantum wells are generally dominated by excitonic contributions and that this is primarily due to the large ratio of the exciton to free carrier dephasing rates. These early reports concluded that DFWM experiments performed with short pulses (~100 fs) always measure signals due to excitons even when more free carriers than excitons are created since the longest dephasing time always dominates. This is the case even if the exciton is only weakly excited by the tail of the pulse. The presence of a large number of potential scattering states facilitate fast dephasing of the continuum states in GaAs QWs and this yields bigger contrast in the relative dephasing rates. In fact, in ref. 26, a 300:1 excitonic resonance was observed.

TABLE II. Calculated absorption strength at $k_x=0$ and their transition energies for 1, 4, and 8 nm $In_xGa_{1-x}N$ QW used in Fig. 4. The spin-orbit splitting of ~10 meV is neglected in order to simplify the presentation of calculation results.

| | Transition label | Normalized absorption strength (a.u.) | Transition energy (eV) |
|---|---|---|---|
| 1 nm QW | $e1$-$h1$, $h2$ | 1, 0.41 | 3.269 |
| 4 nm QW | $e1$-$h1$, $h2$ | 0.05, 0.14 | 3.067 |
| | $e1$-$h3$, $h4$ | 0.31, 0.21 | 3.167 |
| | $e1$-$h6$, $h7$ | 0.44, 0.52 | 3.247 |
| | $e2$-$h1$, $h2$ | 0.33, 1 | 3.251 |
| | $e1$-$h10$ | 0.16 | 3.301 |
| | $e2$-$h3$, $h4$ | 0.18, 0.14 | 3.351 |
| 8 nm QW | $e1$-$h1$, $h2$ | 0, 0 | 2.989 |
| | $e1$-$h5$, $h6$ | 0.06, 0.06 | 3.097 |
| | $e1$-$h8$, $h9$ | 0.13, 0.23 | 3.140 |
| | $e1$-$h10$, $h11$ | 0.37, 0.35 | 3.180 |
| | $e1$-$h12$, $h13$ | 0.49, 0.48 | 3.216 |
| | $e2$-$h1$, $h2$ | 0.04, 0.14 | 3.102 |
| | $e2$-$h3$, $h4$ | 0.29, 0.24 | 3.162 |
| | $e2$-$h5$, $h6$ | 0.47, 0.45 | 3.210 |
| | $e2$-$h8$, $h9$ | 0.23, 0.44 | 3.253 |
| | $e2$-$h10$, $h11$ | 0.07, 0.07 | 3.292 |
| | $e3$-$h1$, $h2$ | 0.25, 0.89 | 3.193 |
| | $e3$-$h8$, $h9$ | 0.07, 0.15 | 3.252 |
| | $e3$-$h10$, $h11$ | 0.17, 0.18 | 3.383 |
| | $e4$-$h1$, $h2$ | 0.26, 1 | 3.275 |
| | $e4$-$h5$, $h6$ | 0.14, 0.15 | 3.384 |
| | $e4$-$h8$, $h9$ | 0.07, 0.14 | 3.427 |
| | $e4$-$h10$, $h11$ | 0.12, 0.13 | 3.466 |
| | $e4$-$h12$, $h13$ | 0.06, 0.08 | 3.502 |
| | $e5$-$h3$, $h4$ | 0.1, 0.09 | 3.375 |

In our case the DFWM experiment picks out localised states with a longer dephasing time from 'less localised' states. Thus, contrary to GaAs, InGaN MQWs DFWM signals represent the selective excitation of more heavily localised states from the distribution that is available, and discriminates against the broad background of 'continuum' states with shorter dephasing times.

## IV. CONCLUSIONS

To summarise, we have performed two-pulse DFWM experiments on InGaN/GaN MQWs which reveal pulse length limited, transient signals. We have observed a resonant enhancement of the DFWM signal at specific wavelengths. Calculations confirm that these signals correspond to excitonic transitions within the quantum wells. The spectral dependence of DFWM provides a novel way to investigate excited state levels in nitride quantum wells with discrimination against the free carrier continuum, information that cannot be obtained using PLE from the highly inhomogeneous InGaN MQWs.

**ACKNOWLEDGEMENTS**
This work has been supported by the EPSRC through grants GR/S24251/01 and GR/R84955/01.